\documentclass[12pt,preprint]{aastex}

\begin{document}

\title {Asymmetry of $^{56}$Ni ejecta and polarization in 
type IIP supernova 2004dj} 

\bigskip

\author{N.N.~Chugai
}
\affil{Institute of astronomy RAS, Moscow 119017, Pyatnitskaya 48}

\begin{abstract}

I study a problem, whether the asymmetry of a $^{56}$Ni 
ejecta that results in the asymmetry of the 
H$\alpha$ emission line at the nebular epoch of the type IIP supernova 
SN~2004dj is able also to account for the recently detected polarization 
of the supernova radiation. I developed a model of the 
H$\alpha$ profile and luminosity with a nonthermal ionization and 
excitation taken into account adopting an asymmetric bipolar 
$^{56}$Ni distribution. On the background of the recovered 
distribution of the electron density I calculated 
the polarized radiation transfer.
It is demonstrated that the observed polarization is reproduced 
at the nebular epoch around day 140 for the same parameters 
of the envelope and $^{56}$Ni distribution for which 
the luminosity and profile of H$\alpha$ are explained. 
Yet the model polarization decreases slower compared to observations.
The origin of an additional component responsible for the 
early polarization on day 107 is discussed. 

\end{abstract}

\newpage

\section{Introduction}

A problem of an explosion mechanism for type II supernovae remains 
unresolved, although this phenomenon 
is undoubtedly related with the gravitational collapse of an iron core.
In this regard of great importance are the observational manifestations 
of the explosion 
that might hint the ways of the problem solution and to play a role of 
observational tests of models. One of the specific properties of 
the explosion mechanism is the asymmetry (Ardeljan et al. 2005;
Blinnikov et al. 1990; Herant et al. 1992; Scheck et al. 2004; 
Burrows et al. 2005). It can manifest itself both in the asymmetry 
of the emission line profiles and in the polarization of radiation. 
Asymmetry effects were observed for the first time 
in the anomalous type IIP supernova SN~1987A (for H$\alpha$ asymmetry 
see Phillips and Williams [1991] and for the polarization 
see Jeffery [1991]). Both effects have been 
interpreted in a model of asymmetric $^{56}$Ni ejecta embedded 
in the symmetric envelope 
(Chugai 1991, 1992). The line asymmetry was observed also in 
the normal type IIP supernova SN~1999em (Elmhamdi et al. 2003). 
Interestingly, in both supernovae the asymmetry was characterized by 
the redshift which indicated that the $^{56}$Ni was ejected 
predominantly toward the far hemisphere.
 
A standard type IIP supernova SN~2004dj (Nakano et al. 2004) 
generated a great interest because of its proximity (D=3.13 Mpc). 
However, more important is that its nebular spectra revealed a strong 
line asymmetry, particularly of the H$\alpha$ emission. 
The asymmetry manifested itself as 
a pronounced blueshift of emission line maximuma by about 1500 km s$^{-1}$ 
(Chugai et al. 2005). Modeling the H$\alpha$ in that paper led 
us to conclude that $^{56}$Ni is distributed as bipolar asymmetric 
ejecta with a more masive jet residing in the near hemisphere.

The recovery of the $^{56}$Ni distribution upon the basis of the 
profile modeling is, strickly speaking, ill-posed problem, so 
doubts may arise whether this procedure is unambiguous.
An independent confirmation is therefore needed. Recently 
Leonard et al. (2006) reported results of the polarimetric 
observations of SN~2004dj in which the intrinsic variable 
polarization in the band of $6800-8200$ \AA\ was detected 
with maximum value 
of 0.56\% at the end of the light curve plateau. 
According to authors suggestion the 
polarization in SN~2004dj arises because of the Thomson scattering, while 
the evolution of the polarization is interpreted as a result of the decrease 
of the Thomson optical depth of the supernovae with spherical envelope 
and asymmetric core. It is shown, simultaneously, that the change of 
position angle of the polarization contradicts to the model 
of axialy symmetric bipolar $^{56}$Ni ejecta.

A question arises whether the asymmetry of the $^{56}$Ni distribution 
responsible for the H$\alpha$ profile asymmetry of the SN~2004dj
is able also to account for the polarization at least 
 at the nebular epoch?
The present paper is concentrated on the answer to this important 
question. In second section I describe the model used 
to compute the electron density distribution caused by the 
asymmetric $^{56}$Ni distribution. The third section presents 
results of the modelling of H$\alpha$ line and of the polarization 
produced by the Thomson scattering of continuum photons in the 
asymmetric electron distribution.

I adopt 2004 June 28 to be the explosion date in 
contrast to June 13 adopted in the previous paper. The revision 
is dictated by two arguments. First, plateau length of 
SN~1999gi, which was used as a template for SN~2004dj, is 
maximum ($\approx125$ d) among type IIP supernovae, while the average 
duration is $\approx110$ d. Second, from 
the spectral evolution of SN~2004dj Patat et al. (2004) estimate 
the explosion date to be about July 14.
Adopting the plateau duration to be 110 d we come to a compromise 
explosion date of June 28, which is two days earlier than the 
explosion date estimated by Vinko et al. (2006) from the analysis 
of the photosphere evolution.

\section{Model of SN~2004dj asymmetry}

The calculation of the intrinsic polarization of SN~2004dj at the 
nebular epoch is reduced to the calculation of (a) the deposition 
of the gamma-rays of $^{56}$Co--$^{56}$Fe decay; (b) ionization 
balance of the hydrogen for the adopted spherical density distribution 
and asymmetric $^{56}$Ni distribution; (c) transfer of the polarized 
radiation on the background of the asymmetric distribution of the
electron density. 
The optimal parameter choice of the $^{56}$Ni distribution is 
determined from the description of 
the H$\alpha$ profile and luminosity. It should be emphasised that 
unlike the previous model, in which the H$\alpha$ emissivity 
was assumed to be proportional to the deposition, here I 
calculate the hydrogen ionization and H$\alpha$ emissivity in 
more detailes. 

The adopted model of the SN~2004dj envelope and $^{56}$Ni distribution,
repeats, with minor exceptions, the model used previously 
 for the H$\alpha$ profile calculation (Chugai et al. 2005).
Specifically, the $^{56}$Ni distribution is represented by the 
central component and bipolar jets. Note, here the central 
$^{56}$Ni component is a complete sphere (Fig. 1), not cut 
sphere as previously assumed (Chugai et al. 2005).
The envelope expands homologously, i.e., the velocity, radius and age 
obey the relation $v=r/t$. The envelope is spherically-symmetric 
(with the exception of $^{56}$Ni) and the density-velocity 
dependence is exponential $\rho=\rho_0\exp(-v/v_0)$, where 
$\rho_0\propto t^{-3}$ and $v_0$ are determined by the 
kinetic energy $E$ and mass $M$. The exponential law qualitatively 
reproduces the density distribution of hydrodynamical models in the 
relevant velocity range $v\leq5000$ km s$^{-1}$. 
The hydrogen abundance in the envelope is $X=0.7$. In the central 
zone $v<v_{\rm s}$, which presumably coincides 
with the spherical $^{56}$Ni component, the 
hydrogen resides, presumably, in condensations with $X=0.7$ and 
the volume filling factor $f<1$.
Note, in the previous paper the parameter $f$ was absent; instead 
it was assumed that $f$ linearly increased between zero and unity 
in the velocity range $0<v<v_{\rm s}$. Following the previus 
paper I assume here that $^{56}$Ni condensations are imbedded 
in the hydrogen-free cocoons with the total mass $M_{\rm c}$. 
The justification and details of the model of the $^{56}$Ni 
distribution along with the method of the calculation of 
the gamma-ray deposition 
are given in the previous paper (Chugai et al. 2005). 

The hydrogen ionization is calculated from equations of the ionization 
balance for two-level hydrogen atom with the continuum. 
In this approximation we are able to take into account major 
processes of the nonthermal ionization and excitation along with the 
photoionization due to the absorption 
of the two-photon and recombination Balmer continuum 
radiation from the second level 
(Chugai 1987; Xu et al. 1992). The radiation transfer in the 
Balmer continuum is treated using  a local approximation.
The absorption probability of the recombination Balmer continuum 
is set to be $w_2=\tau_2/(1+\tau_2)$. Here $\tau_2$ is the local optical 
depth parameter 
$\tau_2=\sigma_2n_2v_0t$, where $\sigma_2$ is the photoionization 
cross-section at the threshold, $n_2$ is the population at the second 
level, $v_0$ is the velocity scale of the exponential density distribution. 
The absorption probability of the two-photon radiation
is approximately $w_{\rm 2q}=\tau_2/(3+\tau_2)$ given 
the frequency dependence of the cross-section ($\sigma\propto\nu^{-3}$) 
and the weak frequency dependence of the two-photon spectrum.

The total rate of the hydrogen nonthermal ionization and excitation is 
\begin{equation}
G=\epsilon(1-\eta)\chi^{-1}\,,
\label{eq:gdef}
\end{equation}
where $\epsilon$ (erg cm$^{-3}$ s$^{-1}$) is the energy deposition 
rate, $\chi$ is the hydrogen ionization potential, 
$\eta$ is the energy fraction spent on the Coulomb heating, which, 
using calculations by Kozma snd Fransson (1992), 
can be approximated as $\eta=x^{0.24}$ (where $x$ is the hydrogen ionization 
fraction). Strickly 
speaking, one needs to take into account that the deposition is shared between
hydrogen and helium. However, since the ultraviolet 
photons emitted by helium owing to the nonthermal ionization and excitation 
efficiently ionize hydrogen, the omission of 
the nonthermal ionization of He does not markedly change the rate of 
hydrogen ionization and excitation compared with
the approximation (\ref{eq:gdef}).

I assume following Xu et al. (1992) that the energy spent on the 
nonthermal hydrogen ionization and excitation is shared between 
the ionization with the branching ratio $\phi_1=0.39$, second level excitation 
($\phi_2=0.47$) and third level excitation ($\phi_3=0.14$). Collisions 
with thermal electron are also taken into account. Since a thermal balance 
is not calculated I adopt the constant electron temperature 
$T_e=5000$~K. The model with $T_e=4500$~K produces 
only 5\% lower emission measure, i.e., $\sim2.5$\% lower electron 
concentration. This demonstrates a weak sensitivity to 
the electron temperature in the region of reasonable values of this 
paprameter.
Balance equations for the ionization and second level population with 
principal processes taken into account read 
\begin{equation}
G\phi_1+n_2(P_2+q_2n_e)+q_1n_en_1=\alpha_{\rm B}n_e^2\,,
\label{eq:g1}
\end{equation}
\begin{equation}
n_2(A_{\rm2q}+q_{21}n_e+A_{21}\beta_{12}+P_2+q_2ne)=\alpha_{\rm B}n_e^2+
G(1-\phi_1)+q_{12}n_en_1\,,
\label{eq:g2}
\end{equation}
where $P_2$ is the photoionization rate for the second level, 
$\alpha_{\rm B}$ is the recombination coefficient for excited 
levels $n\geq2$, $n_e$ is the electron concentration, 
$A_{\rm 2q}=2.06$ s$^{-1}$ is the probability of the two-photon 
transition 2--1 (I assume the equipartition between 
$2s$ and $2p$ levels), $q_{12}$, $q_{21}$ are the 
collisional excitation coefficients,
$q_1$, $q_2$ are collisional ionization coefficients, 
$A_{21}$ is the spontaneous transition rate, 
$\beta_{12}$ is the Sobolev escape probability fot the L$\alpha$ photon.
In equation (\ref{eq:g2}) I adopt that the nonthermal
 excitation of the third level (second term in the right hand side) 
results in the excitation of the second level. The photoionization 
rate is determined by the absorption of the two-photon and recombination 
Balmer continuum radiation: 
\begin{equation}
n_2P_2=1.42n_2A_{\rm 2q}w_{\rm 2q}+\alpha_2n_e^2w_2\,.
\label{eq:ph2}
\end{equation}
The factor 1.42 takes into account the fraction (0.71) of the 
two-photon radiation with energy $>3.4$ eV and that two photons 
are emitted simultaneously. 

Optimal parameters of the $^{56}$Ni distribution are choosen via 
modelling the H$\alpha$ profile and its evolution. The 
H$\alpha$ emissivity is calculated using the balance equation 
for the third level that includes the major processes: 
recombinations to levels $n\geq3$, nonthermal excitation of the 
third level by fast electrons and de-excitation by thermal
electron collisions. 
The net rate of the H$\alpha$ emission (erg cm$^{-3}$ s$^{-1}$) is then
\begin{equation}
\epsilon_{32}=h\nu_{23}(\alpha_{\rm C}n_e^2+Gf_3+q_{23}n_en_2+q_{13}n_en_1)
\frac{A_{32}\beta_{23}}{A_{32}\beta_{23}+q_{32}n_e}\,,
\end{equation}
where $\alpha_{\rm C}$ is the recombination coefficient 
to levels $n\geq3$. The above equations of ionization and level populations 
balance are solved for the envelope with the kinetic energy 
of $E=1.5\times10^{51}$ erg that is characteristic of SN~1987A (Utrobin 2005)
and ejecta mass $M=15~M_{\odot}$. The adopted $^{56}$Ni mass is 
$0.02~M_{\odot}$ (Chugai et al. 2005). Our computations show 
a weak dependence on the energy. The model with lower energy, 
$E=10^{51}$ erg being characteristic of SN~1999em (Baklanov et al. 2005) 
predicts practically the same H$\alpha$ profile and the very 
small difference in the H$\alpha$ absolute luminosity behavior. 

The transfer of the polarized radiation of the quasi-continuum in the 
$7000-8000$~\AA\ band is modelled by Monte Carlo technique 
(Angel 1969). I consider only Thomson scattering of photons. 
The direction and polarization 
of the scattered photon is diced according to the dipole scattering law.
Stokes vector components $I$, $Q$, and $U$ of escaping photons 
are summed in the corresponding 
polar angle bin. The number of photons in a typical simulation 
is $\sim10^7$.
The code was tested using available analytical results. In particular, 
the model reproduces the Chandrasekhar-Sobolev limit ($p=11.7$\%) 
in the problem of the polarization for the Thomson scattering in the plane 
atmosphere.

\section{Results}

Calculations of the H$\alpha$ profile and polarization 
were performed for the models with the inclination angle of the $^{56}$Ni 
ejecta $i=30^{\circ}$ (model M1), likewise in the previous model 
(Chugai et al. 2005), and $i=40^{\circ}$ (model M2). Parameters of 
the $^{56}$Ni distribution for all the models, including 
previous model M0 (Chugai et al. 2005), are given in Table 1.
Starting with the third column the Table presents the expansion velocity 
of the central component, boundary velocities of jets in the near 
and far hemisphere, mass of the central $^{56}$Ni component, 
masses of both $^{56}$Ni jets, total mass of cocoons, and the volume 
filling factor $f$ of the hydrogen in the central zone (in the model M0 
this parameter is absent). Parameters of the model M1 are a bit different   
from those of the model M0 because the model M1 has a different 
geometry of the central component, the model M1 takes into account 
the filling factor, and the model 
describes sources of H$\alpha$ photons in greater details.
  
Both M1 and M2 models sensibly reproduce the observed H$\alpha$ 
luminosity (Fig. 2). This provides a confidence that 
the model includes essential processes of the hydrogen 
ionization and excitation in the envelope. Additional support for the 
model comes from a close similarity between 
calculated and observed H$\alpha$ profiles for different moments 
(Fig. 3). This confirms the major result of the 
previous simulations performed in the frame of a more simple model M0. 
one should emphasise, however, that the absolute correspondence between 
model and observed profile is lacking. This indicates that the real 
supernova has, probably, more complicated structure than our model. 

The polarization caused by the Thomson scattering is computed using
the distribution of the electron concentration found from 
the modelling of the H$\alpha$ profile. Isodensity lines for the 
electron concentration in the model M1 
on days 116 and 315 are shown in Fig. 4. The hourglass structure on day 
116 is due to the small (compared with radius)
free path length of gamma-quanta in the inner 
region and large path length in the outer layers. 
With time the $n_e$ distribution gets more spherical because of the 
increase of the free path length of gamma-quanta.
To compute the polarization we must also set the distribution of 
the photon sources. I follow a reasonable assumption that the emissivity of 
the quasicontinuum is proportional to the deposition rate $\epsilon$. 

The polarization in model M1 and M2 on days 116 and 222 
is shown in Fig. 5 as a function 
of cosine of the polar angle $\theta$ measured from the axis of 
the rotational symmetry. Both models result in the similar 
polarization for the same angles. However, for specific inclination 
angles of the models, i.e., $i=30^{\circ}$ for M1 and $i=40^{\circ}$ 
for M2, the polarization in the model M2 is significantly (1.5--2 times) 
larger than in the model M1. The polarization decline with time 
reflects both the decrease of the Thomson optical depth and 
spherization of the distribution of the electron concentration. 
Note, the position angle of the polarization vector in our model is 
perpendicular to the axis of the rotational symmetry.

It is instructive to calculate the polarization 
in the model M1Ni (Fig. 5) which differ from the model M1 by the 
larger $^{56}$Ni mass ($0.075~M_{\odot}$ versus $0.02~M_{\odot}$). 
The $^{56}$Ni mass in the model M1Ni is equal to that found 
in SN~1987A. This calculation demonstrates the polarization 
for a hypothetical SN~IIP with the $^{56}$Ni mass to be characteristic of 
SN~1987A and $^{56}$Ni asymmetry to be characteristic of SN~2004dj.  
Interestingly, the polarization for the inclination angle 
$i\sim90^{\circ}$ in the model M1Ni is only by a factor of 1.44 larger than 
in the model M1. This means that the dependence of the polarization on 
the $^{56}$Ni mass is somewhat weaker than square root of the mass. 
The latter is expected in a naive model with the recombination 
rate to be proportional to the deposition rate. 
Let us compare the computed polarization in the 
model M1Ni with the observed polarization of SN~1987A at the 
nebular epoch. Around day 200 the intrinsic
polarization of SN~1987A in $R$ band (where effect of line scattering 
is small) is about $0.9\pm0.17$\% (cf. Jeffery 1991). According 
to the model M1Ni this value corresponds to the inclination angle 
of the bipolar ejecta $i\sim53\pm10^{\circ}$.
This estimate should be regarded, of course, as illustrative. Yet, 
it is amazing that the value coincides within errors with the 
inclination of the bipolar ejecta of SN~1987A $i\approx45^{\circ}$ 
estimated from different arguments (Wang et al. 2002).

I return now to the polarization evolution 
in SN~2004dj. The calculated polarization in models M1 and M2 
taking into account their inclination angles (Table 1) is 
shown in Fig. 6 together with the observed polarization for 
SN~2004dj reported by Leonard et al. (2006).
The model M2 predicts too large polarization compared to observations 
and thus should be 
discarded. The polarization in the model M1 reproduces the 
observed value on day 144. However, the later decline of the model
polarization is slower compared to the observed polarization. 
This mismatch possibly stems from the model drawback 
responsible for poor description of the central part of 
the H$\alpha$ profile on days 222 and 315 (Fig. 3).
Our model probably is not quite adequate in the description of 
the central zone structure, where deviations from the 
bipolar structure are conceivable. Another drawback of the 
model is somewhat lower polarization around day 110 compared 
to observational value (Fig. 6). The inconsistency of the model 
at this epoch is, however, unavoidable by, at least, two reasons.

First, on day 107 the supernova is still at the photospheric stage, 
although at the very end of it. At this epoch the 
thermal ionization is dominant, which makes our model unapplicable.
The role of 
the thermal ionization at that time is illustrated by Fig. 7 which 
shows the H$\alpha$ line on days 107 (Leonard et al. 2006), 
112, and 116 (Chugai et al. 2005). On day 116, as we saw above (Fig. 3), 
the H$\alpha$ line is well described by the model of the nonthermal 
ionization caused by the asymmetric bipolar $^{56}$Ni ejecta.
A bit earlier, on day 112, the profile shows symmetric component 
related with the residual thermal ionization. On day 107 the 
symmetric component related with the thermal ionization is 
dominant (Fig. 7). Our model thus is not applicable for the 
description of the polarization at this stage. 

The second reason, why the model is not valid on day 107, is 
even more crucial. According to polarization data (Leonard et al. 2006) 
the position angle (PA) of the polarization vector is $28^{\circ}$, whereas  
for the next observation momemt (144 d) and later on 
PA$\approx178^{\circ}$. Between days 107 and 144, thus, PA rapidly changes 
by $\approx30^{\circ}$. The persistence of PA at the nebular epoch is 
consistent with the model of the axisymmetric bipolar $^{56}$Ni ejecta, 
while the observed change of PA at the early epoch, on the contrary, 
disagrees with this model. Generally, the transition from 
optically thick to optically thin scattering regime may cause a 
jump of PA of the polarization vector by $90^{\circ}$ (Angel 1969). 
This phenomenon, obviously, has nothing to do with the observed PA
jump by $\approx30^{\circ}$. The behavior of PA between days 104 and 144 
indicates, therefore, that some additional transient 
component of polarized radiation 
dominates around day 107 and this component is not related to axisymmetric  
$^{56}$Ni ejecta (Leonard et al. 2006). The early polarization 
component may be dubbed for clarity
 "photospheric" in contrast to the "nebular" at the epochs $t\geq140$ d.

In the range of the applicability of the axysymmetric model 
($t\geq140$ d) one may claim that the model of the 
bipolar $^{56}$Ni ejecta with the inclination angle $i=30^{\circ}$
that describes the H$\alpha$ profile also reproduces the observed 
polarization at the early nebular epoch 
($t\sim130-150$ d) and is consistent qualitatively with the 
subsequent polarization decrease. It should be emphasised, however, that
at the late nebular epoch ($t\sim 200$ d) the theoretical polarization 
is somewhat higher than the observed one.

\section{Conclusion}

The primary goal of the paper was to provide an answer to the 
question whether the asymmetry of the $^{56}$Ni distribution 
responsible for the asymmetry of the H$\alpha$ emission line at the 
nebular epoch of SN 2004dj 
is able to account for the observed polarization as well.
I constructed a model of 
the H$\alpha$ profile and luminosity with the nonthermal 
excitation and ionization of hydrogen for the asymmetric $^{56}$Ni 
distribution. I then calculated the transfer of a polarized radiation 
using Monte Carlo technique. The modelling shows that the 
model of the bipolar ejecta, which describes the H$\alpha$ line, predicts 
that the model polarization is consistent with the observed polarization 
in SN~2004dj at the early nebular epoch.
At the late stage $t\sim 200$ d  the theoretical polarization is somewhat 
larger compared with observations. This disparity is possibly 
related with a more complicated structure of the matter distribution, 
including $^{56}$Ni, in the inner zone of the envelope. 

The photospheric component of the polarization observed on day 107 is 
not described in principle by the model of the axisymmtric 
$^{56}$Ni ejecta. Leonard et al. (2006) attribute the early 
polarization to the deviation of the bipolar 
$^{56}$Ni ejecta from  the axial symmetry. According to their conjecture 
the recombination wave propagating toward the center of the supernova 
envelope uncovers in sequence layers with different position 
angles of $^{56}$Ni clumps. This might explain the observed change of 
polarization position angle between days 107 and 144. 
In this model the photosphere 
plays a role of a homogeneous screen, while asymmetry arise from the enhanced 
ionization originated from the asymmetric $^{56}$Ni distribution. 
The scenario proposed by Leonard et al. (2006) 
can be dubbed "the model of non-axisymmetric $^{56}$Ni ejecta".

An alternative conjecture that does not require abandoning the axial 
symmetry of $^{56}$Ni ejecta is conceivable either. 
Note, the photospheric component 
of the polarization emerged at the phase when the photosphere almost 
reached the center of the supernova envelope. On the other hand, 
the distribution of the chemical 
 abundance in the inner layers of supernova should be 
 essentially inhomogeneous due to 
an incomplete mixing of the hydrogen and helium enevelopes 
 during the shock wave propagation (M\"{u}ller et al. 1991).
The photosphere that forms in the chemically inhomogeneous material 
unavoidably should acquire large scale brightness variations that 
might result in a significant polarization of the photospheric 
radiation. Such a scenario one may dub a model of "spotty photosphere" 
in contrast to the model of non-axysymmetric $^{56}$Ni ejecta. 

In both conjectures the early polarization is related 
to the final photospheric phase and in both 
mechanisms a "flash" of polarization should generally
emerge at the end of the light curve plateau. The difference is 
that in the model of the spotty photosphere the polarization flash 
should be present even in SN~IIP with the almost spherical $^{56}$Ni
 distribution, while the model based on the $^{56}$Ni asymmetry 
 predicts in this case essentially weak/no polarization.  
This signature can be used to discriminate between alternative 
conjectures about the origin of the photospheric component 
of polarization in SN~2004dj and other SN~IIP with the detected 
polarization flash.

\bigskip 

I am grateful to Douglas Leonard for the kind permission to use 
the spectrum of SN~2004dj. The work is partially supported by RFBR 
grant 04-02-17255.

{}

\clearpage

%======================================================================
\begin{table}
  \caption{Parameters of $^{56}$Ni distribution}
  \bigskip
  \begin{tabular}{lccccccccc}
  \hline

Model & $i$ & $v_{\rm s}$  &  $v_1$ & $v_2$ & $M_{\rm s}$ & $M_1$ & 
 $M_2$  &  $M_{\rm c}$ & $f$ \\
 \hline 
 &   &\multicolumn{3}{c}{km s$^{-1}$} & \multicolumn{4}{c}{$M_{\odot}$}\\

\hline                                         

M1 & $30^{\circ}$ & 1300 & 2600 & 3800 & 0.0078 & 0.009 & 0.0032 & 0.8 & 0.3\\
M2 & $40^{\circ}$ & 1250 & 3100 & 4500 & 0.0086 & 0.0083 & 0.0031 & 0.9 & 0.4\\
M0 & $30^{\circ}$ & 1400 & 2700 & 3500 & 0.0078 & 0.0083 & 0.0038 & 1.3 & \\
\hline
\end{tabular}
\label{t-par}
\end{table}
%=================================================================================

\clearpage

%======================================================================
 \begin{figure}
%\epsscale{.80}
\plotone{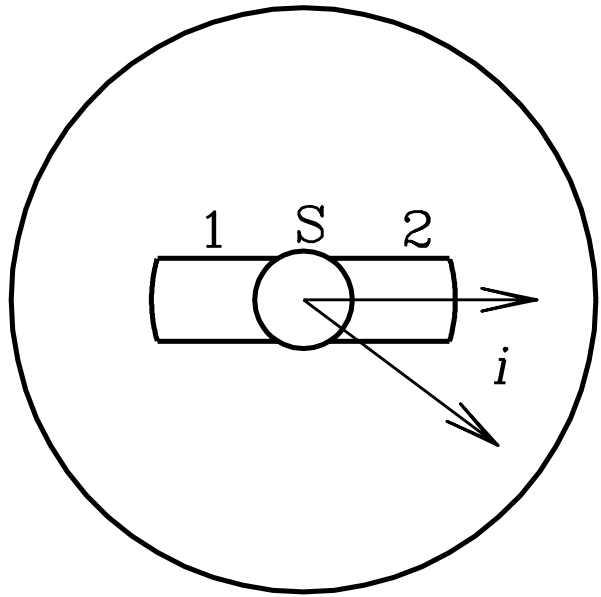}
\caption{Schematic representation of 
the spherical envelope with bipolar $^{56}$Ni ejecta inside. 
Nikel ejecta with the inclination angle $i$ consists of the 
central spherical component S and two cylindrical jets 
(jet 1 is in near hemisphere).
}
\end{figure}
%=====================================================================
\clearpage

%======================================================================
 \begin{figure}
%\epsscale{.80}
\plotone{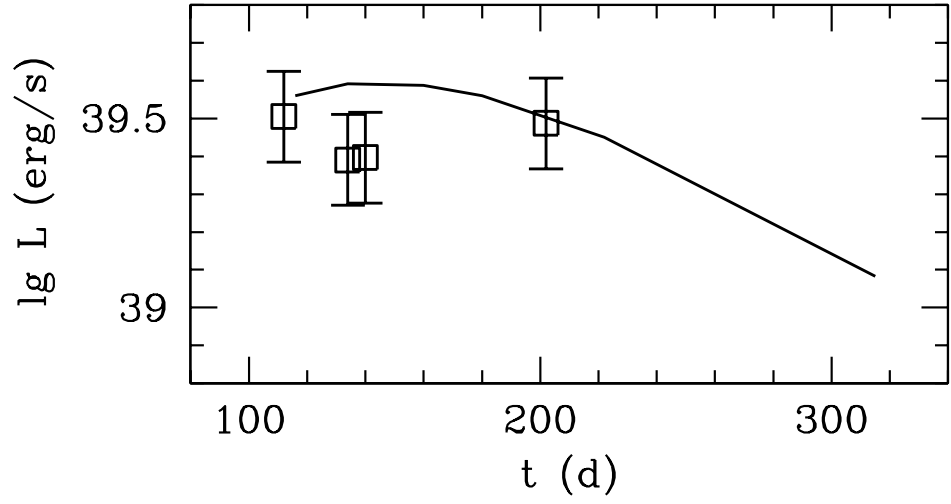}
\caption{H$\alpha$ luminosity of SN 2004dj according to observations 
({\em squares}) and models M1 and M2 ({\em line}).
}
\end{figure}
%=====================================================================
\clearpage

%======================================================================
 \begin{figure}
%\epsscale{.80}
\plotone{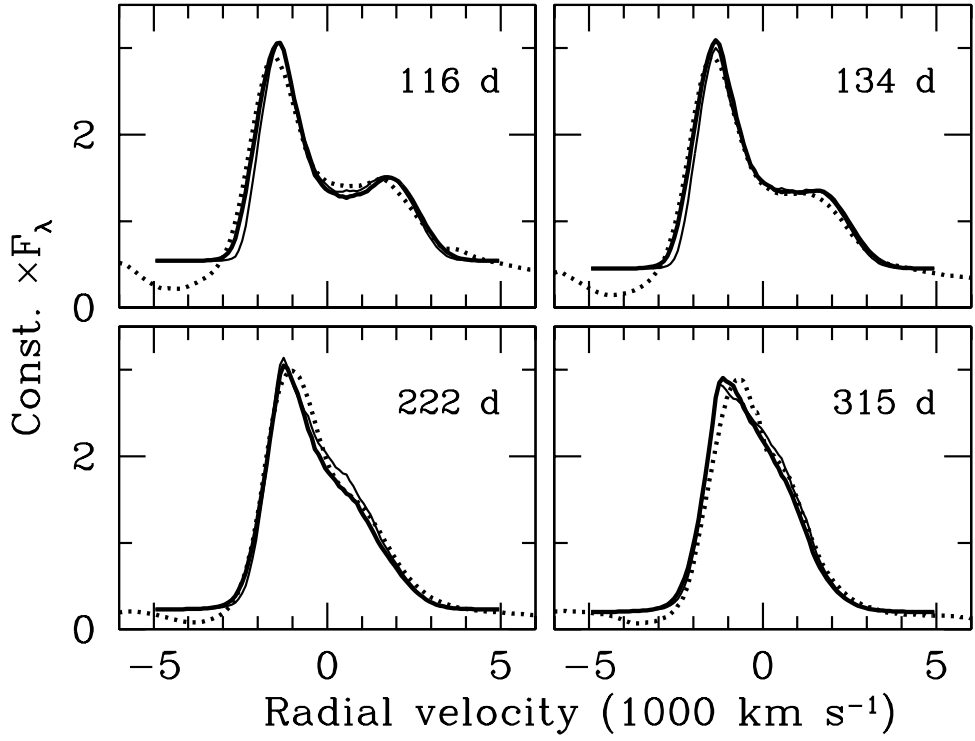}
\caption{H$\alpha$ profile for different moments in the model M1 
({\em thin solid line}) and in the model M2 ({\em thick solid line}) 
as compared to observations ({\em dotted line}). The flux is in 
arbitrary units. 
}
\end{figure}
%=====================================================================
\clearpage

%======================================================================
 \begin{figure}
%\epsscale{.80}
\plotone{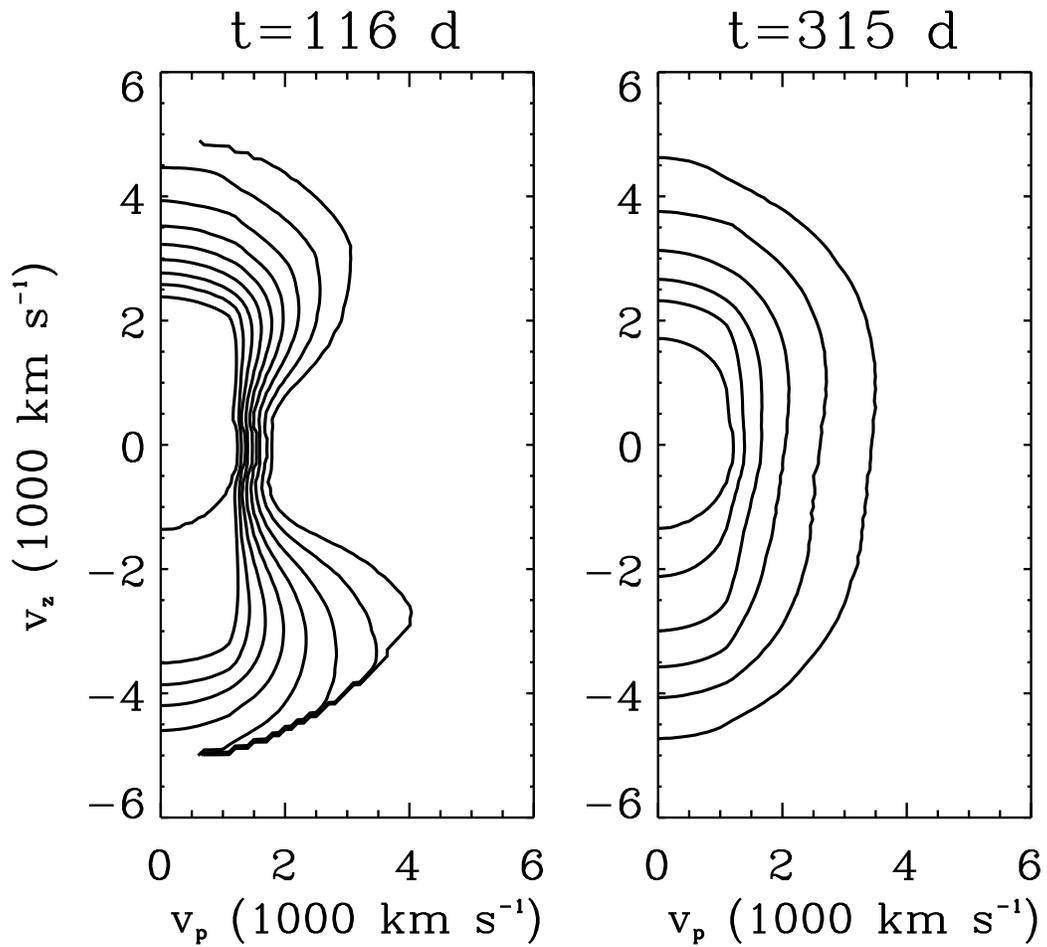}
\caption{Isodensity lines of the electron concentration 
in meridional plane for the model M1 in two phases.
The ordinate is the velocity along the axis of ejecta, 
the absciss is the velocity in a perpendicular direction.
The concentration varies by a factor of two between neighbouring 
contours. The external contour corresponds to the concentration 
$10^6$ cm$^{-3}$.
}
\end{figure}
%=====================================================================
\clearpage

%======================================================================
 \begin{figure}
%\epsscale{.80}
\plotone{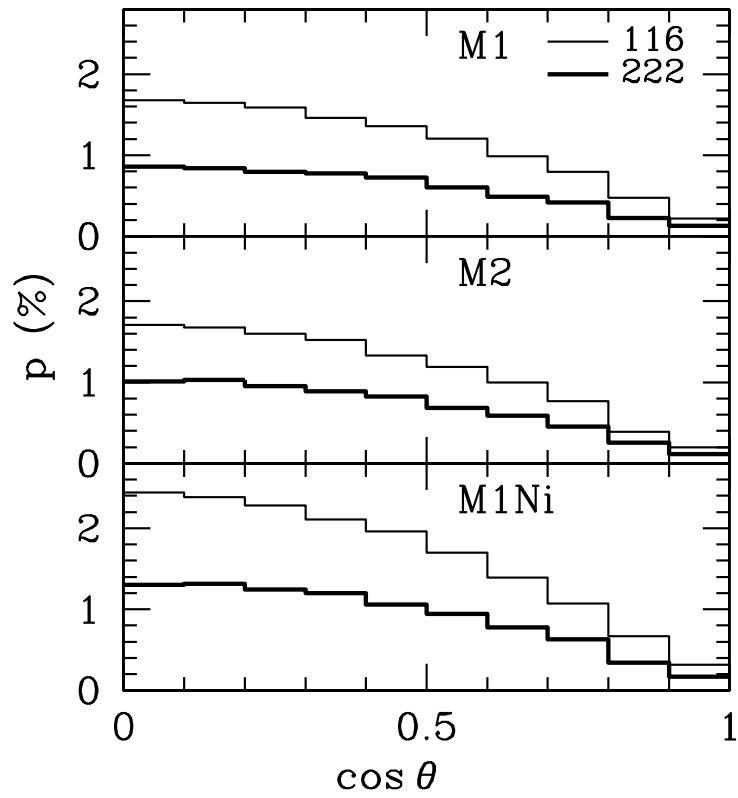}
\caption{Polarization for two moments as a function of  
cosine of the polar angle for the model M1 ({\em upper panel}), 
M2 ({\em middle panel}), and M1Ni ({\em lower panel})
}
\end{figure}
%=====================================================================
\clearpage

%======================================================================
 \begin{figure}
%\epsscale{.80}
\plotone{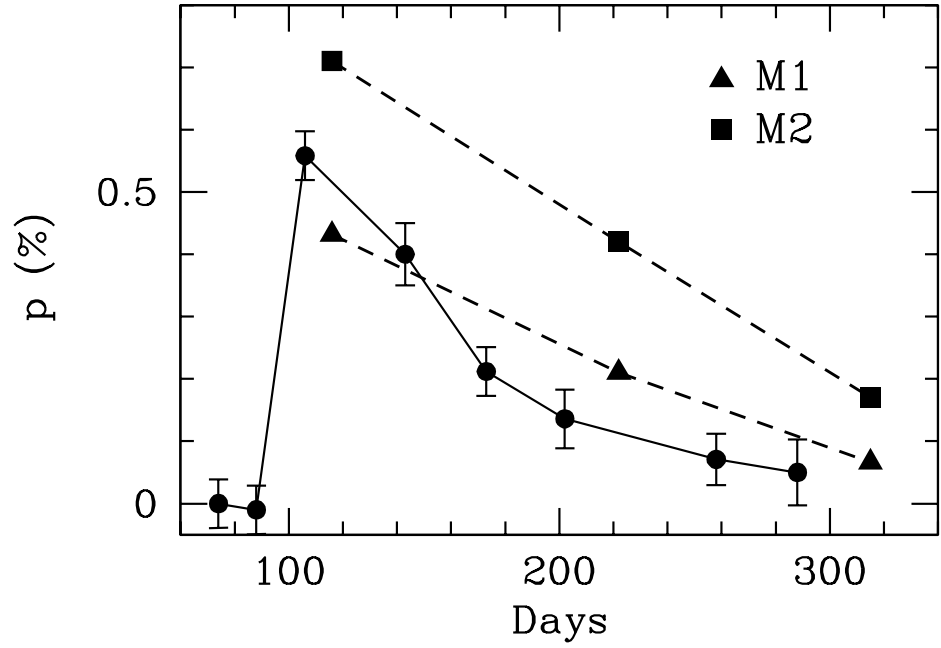}
\caption{ Evolution of polarization in the models M1 and M2
compared with the observed polarization ({\em filled circles}).
}
\end{figure}
%=====================================================================
\clearpage

%======================================================================
 \begin{figure}
%\epsscale{.80}
\plotone{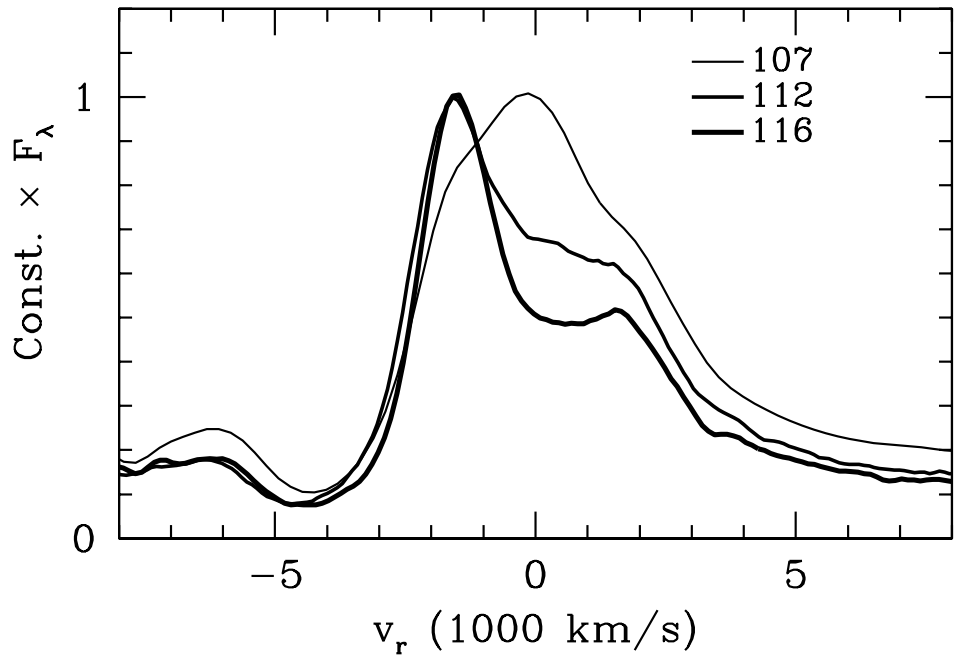}
\caption{H$\alpha$ evolution between days 107 and 116 at the 
phase of the transition from the light curve plateau to the radioactive tail. 
The flux in maxima is normalized on the same value. The symmetric component 
of ionization apparently dominates on day 107 and then quickly fades 
during subsequent 9 days. 
}
\end{figure}
%=====================================================================

\end{document}